\documentclass[10pt]{elsart}

\usepackage{amssymb,amsmath}

\usepackage{graphics,graphicx}
\usepackage{amsmath,amssymb,float}
\begin{document}
\runauthor{Endler and Gallas}
\begin{frontmatter}
\title{Existence and characterization of stable ghost orbits in the H\'enon map}
\author[a1,a2]{Ant\^onio Endler}  
\author[a1,a2]{Jason A.C.~Gallas} 
\address[a1]{Instituto de F\'\i sica, 
             Universidade Federal do Rio Grande do Sul, \\
            91501-970 Porto Alegre, Brazil}
\address[a2]{Institut f\"ur Computer Anwendungen, 
             Universit\"at Stuttgart, \\
             D-70569 Stuttgart, Germany}

\begin{abstract}
We report a remarkable type of bifurcation:
by varying real parameters, unstable complex orbits may become
stable over wide parameter ranges. 
Thus, phase diagrams obtained by analizing solely the stability
of real solutions may be incomplete.
\end{abstract}
\end{frontmatter}

The purpose of this paper is to report a remarkable new type 
of bifurcation: 
unstable complex orbits may be stabilized by varying {\it real\/} 
model parameters. In other words, by varying real parameters
it is possible to stabilize ``complex phases'' in phase-diagrams.
This surprising fact is shown for the paradigmatic example of a 
multidimensional dissipative dynamical system, 
the H\'enon map $(x,y) \mapsto (a - x^2 + by, x)$.
The parameter space of the map contains a {\it wide\/} domain of 
{\it real\/} parameters $a$ and $b$ where it is possible
to find complex `ghost' solutions which are {\it stable}.

This new bifurcation is of importance in the construction of phase 
diagrams, 
usually constructed by sweeping real parameters and studying the
set of real solutions, since domains of complex stable motions might be 
missing in them.
Another interesting implication is that the plethora of
ghost (complex) orbits, so fundamental nowadays in quite
different fields\cite{imp1,teresa,imp4}, 
may be subdivided dichotomically into {\it unstable\/} and {\it stable\/}
ghosts, pointing to the necessity of investigating the effect
of complex stability in all physical applications.
In atomic physics, for instance, the stabilization of complex ghost
orbits is expected to allow sum re-arrangements in
trace formulas\cite{gutz}.
From the exact analytical results reported here one can show that 
the H\'enon map displays Naimark-Sacker bifurcations and, consequently, 
supports quasiperiodic behaviors\cite{enga04}.

The possibility of stabilizing complex orbits seems not to have 
been considered before\cite{greenfield,biham}, perhaps because the
algebraic varieties involved are of very high degrees, 
exceeding by far those studied by mathematicians\cite{liv1,liv2}. 
The stabilization of complex orbits was not considered in the 
classic work of Arnold\cite{arnold}.

As shown recently\cite{eg2002,eg2001},
one may always reduce the equations of motion of any algebraic 
dynamical systems to a pair of polynomial equations:
(i) ${\mathbb P}(x,\sigma)=0$, parameterizing simultaneously 
    {\it all orbits of any given period\/} in terms of the sum 
    $\sigma$ of orbital points, and
(ii) ${\mathbb S}(\sigma)=0$, defining the values
     of $\sigma$ as a function of model parameters.
The degree of ${\mathbb S}(\sigma)$  tells the quantity of
independent solutions available.
In addition, it is also possible to write the secular equation 
ruling the stability of the system as a function of $\sigma$ 
and of model parameters.
Following Ref.~\cite{eg2002}, the polynomials
providing {\it complete\/} information about all 
possible period-$6$ orbits are
\begin{eqnarray}
{\mathbb P}(x,\sigma) &=&   x^6 - \sigma x^5 
                          + \theta_4(\sigma) x^4  
                          - \theta_3(\sigma) x^3 
                          + \theta_2(\sigma) x^2 
                          - \theta_1(\sigma) x   
                          + \theta_0(\sigma),    
                                                 \label{eqorb}  \\
{\mathbb S}(\sigma) &=& \sum\limits_{i=0}^{9}\ \Theta_i(a,b)\sigma^i,       
                                                    \label{eqsigma} 
\end{eqnarray} 
where $\sigma \equiv x_1 + x_2 + x_3 + x_4 + x_5 + x_6$ is the sum of the
orbital points. 
The coefficients  $\theta_i(\sigma) \equiv \theta_i(a,b;\sigma)$  
are the standard symmetric functions of the roots $x_\ell$ (orbital points).
The coefficients $\theta_i(\sigma)$ and $\Theta_i(a,b)$ are given 
explicitly in the  Appendix at the end of the manuscript. 
Additionally, orbital stability is ruled by the following quadratic
for the eigenvalues $\lambda$:
\begin{equation}
{\mathbb L}(\lambda) = \lambda^2 
    -\big[ 2b^3 + \big( \frac{ {\mathcal N}_1 + 2{\mathcal N }_4}
              { {\mathcal D }}\big) b^2  
- 2\frac{{\mathcal N}_5}{ {\mathcal D } }b + 64\theta_0\big]\lambda  + b^6 
     \label{eqeigen}
\end{equation}
where ${\mathcal N}_5$, not contained in
Eqs.~(\ref{eqorb})-(\ref{eqsigma}), is given in the Appendix. 
Equations (\ref{eqorb})-(\ref{eqeigen}) result from quite long 
algebraic manipulations which were performed automatically on a 
computer using specially devised {\it ad hoc\/} routines.

The degree of Eq.~(\ref{eqsigma}) tells that for any set $(a,b)$
of parameters there are nine possible period-$6$ orbits, 
not all necessarily different. 
The actual  orbits are found by substituting the nine roots of 
Eq.~(\ref{eqsigma}) into Eq.~(\ref{eqorb}).
Since for real parameters all $\Theta_i(a,b)$ are real, 
(i) there is always at least one real value of $\sigma$, and
(ii) complex values of $\sigma$ must always appear in {\it conjugate pairs}.

\begin{figure}
\includegraphics[width=14.0cm]{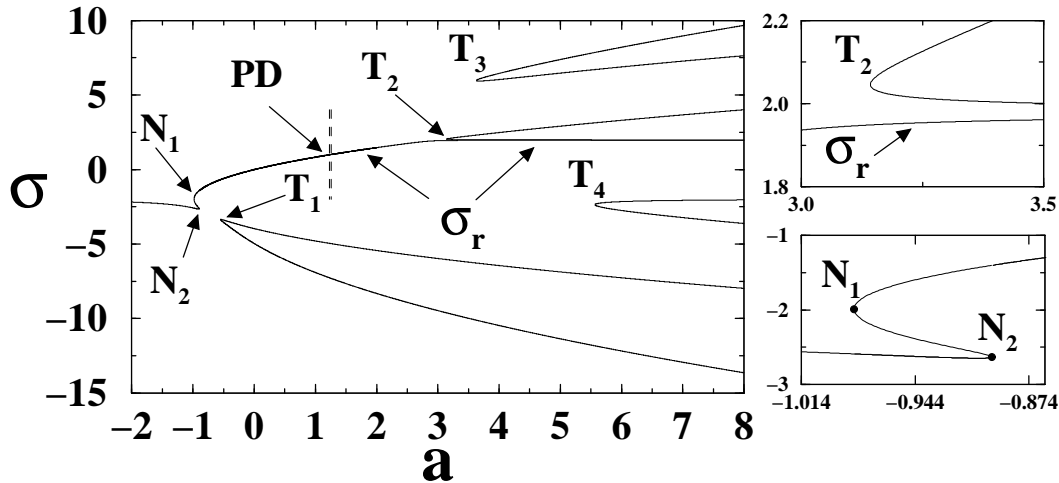}
\caption{\label{fig:fig1}  
  Real roots of ${\mathbb S}(\sigma)=0$ as a function of $a$, for $b=-0.98$. 
  Stable complex orbits exist between $N_1$ and $N_2$ on the locus
  $\sigma_R$, the locus of the real root that is always present.
  PD refers to the period-doubling, $T_i$ to tangent bifurcations.
  The figures on the right show  details hard to see on 
  the left. See text.}
\end{figure}

\begin{figure}[thb]
\begin{minipage}[t]{65mm}
\includegraphics[width=7.4cm]{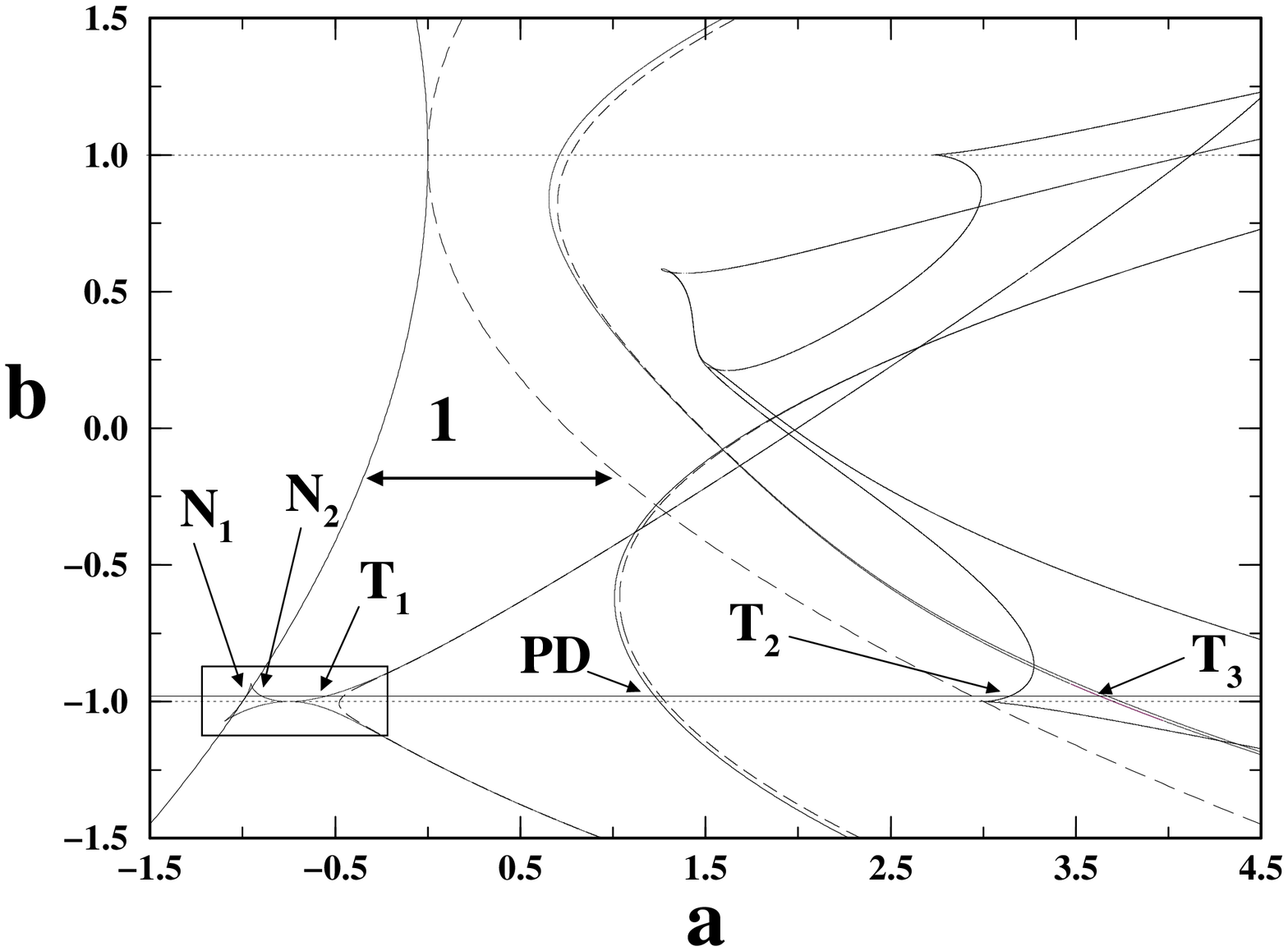}
\caption{Stability domains and singularities for
         period-$6$ motions. See text.}
\label{fig:fig2}
\end{minipage}
\hspace{\fill}
\begin{minipage}[t]{65mm}
\includegraphics[width=6.4cm]{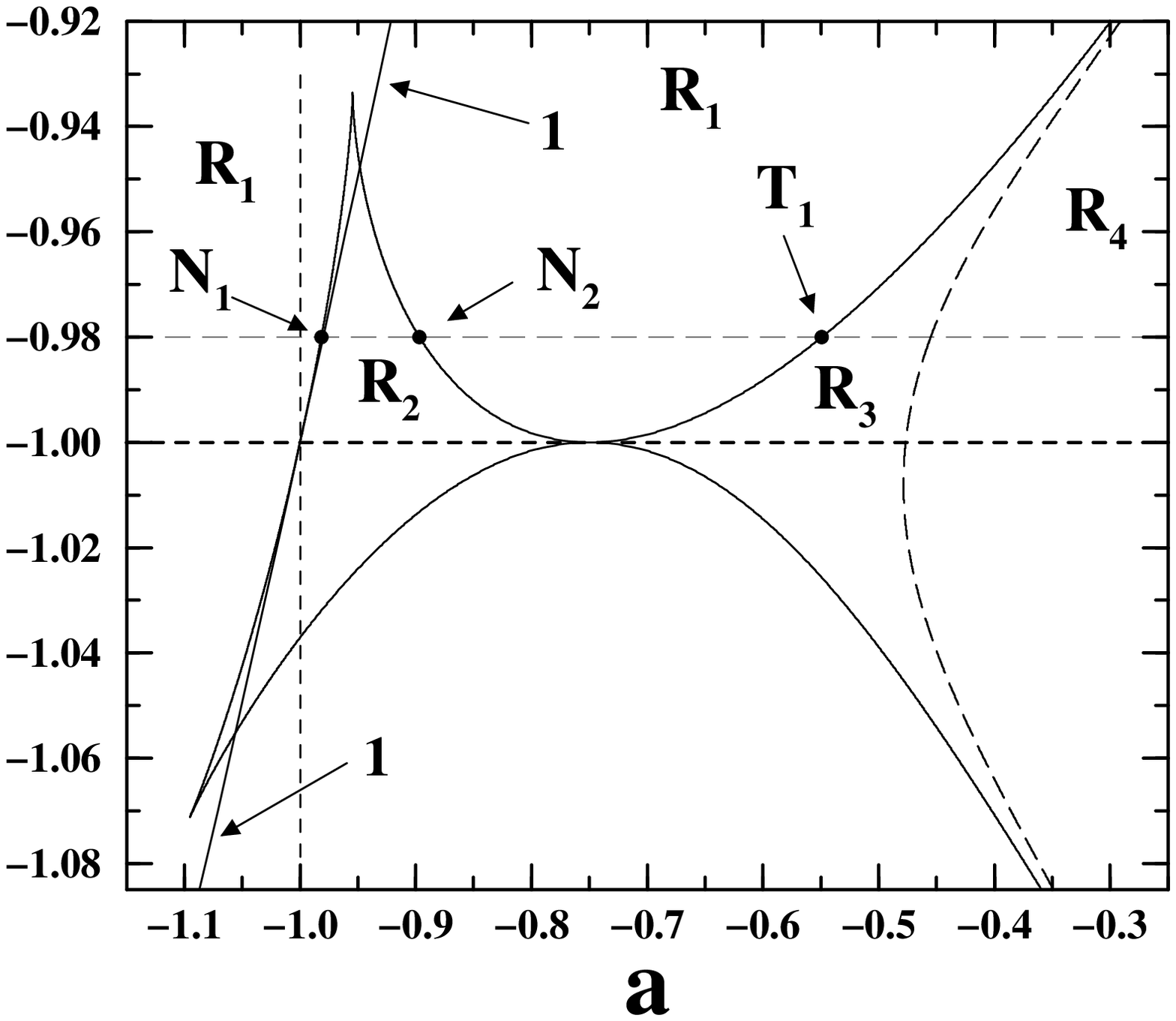}
\caption{Blow-up of the box in Fig.~\ref{fig:fig2}.
       Stable complex orbits exists in $R_2$.}
\label{fig:fig3}
\end{minipage}
\end{figure}

To get a feeling about the nature of the foliated surface defining $\sigma$ 
values, Fig.~(\ref{fig:fig1}) shows the real roots of 
${\mathbb S}(\sigma)=0$ as a function of $a$,  for $b=-0.98$. 
The four points $T_i$ indicate the location of tangent bifurcations.
Of special interest is the locus $\sigma_R$ defined by that root
of Eq.~(\ref{eqsigma}) that is always real.
Along this locus two different phenomena occur. 
First,  one finds the familiar  $3\to6$ period-doubling bifurcation, 
indicated by PD.
The doubled orbit is stable in a small interval between two vertical
dashed lines that is too small to be discernible in the figure.
Second,  along $\sigma_R$ we find a remarkable
{\it new type of bifurcation\/} arising from the {\it multivalued\/} 
character of $\sigma_R$ between the points $N_1$ and $N_2$.
In this interval there are three real roots $\sigma$,
which define three {\it complex\/} orbits.
The complex orbit defined by the ``middle branch'' interconnecting
$N_1$ and $N_2$, is {\it stable} inside a region
resembling a ``bow-tie'' [see Fig.~(\ref{fig:fig3}) below]
precisely where $\sigma_R$ displays a fold 
[in the interval between $N_1$ and $N_2$ in Fig.~(\ref{fig:fig1})]
This shows that {\it folds are not always necessarily connected only 
with tangent bifurcations}.

Full lines in
Fig.~(\ref{fig:fig2}) show how the singularities in Fig.~(\ref{fig:fig1})
evolve when $b$ changes.
Dotted lines, obtained by investigating eigenvalues, delimit stability 
domains in the usual way. For reference, Fig.~(\ref{fig:fig2})
also displays the interval where period-$1$ orbits (fixed points) are stable.
The new bifurcation being reported here occurs inside the box, shown 
magnified in Fig.~(\ref{fig:fig3}).
As is known, tangent bifurcations may be located analytically from the
discriminant of Eq.~(\ref{eqsigma}) with respect to $\sigma$.

Figure (\ref{fig:fig3}) shows the stability domain of complex orbits
along with a line ``1'' marking the border where stable orbits of
period-$1$ are born when $a$ increases.
Recall that physically meaningful solutions exist only in the interval
$-1 \leq b \leq 1$. 
The scenario in the region above the line $b=-1$ is divided into four 
domains labeled $R_i$ and displays the following characteristics.
All nine period-$6$ orbits are complex in $R_1$, despite the fact that
one of them corresponds to a real $\sigma$.
Moving from $R_1$ into $R_2$, one finds the complex orbit stabilizing 
bifurcation, when two additional 
values of $\sigma$ become real (see curve $\sigma_R$ in Fig.~(\ref{fig:fig1})) 
but their corresponding orbits remain complex, one of them being stable.
In $R_3$ there are three orbits associated with real values of $\sigma$,
two orbits being real, one stable and one unstable. 
The complex orbit is unstable, being the same orbit that will give rise
to a stable orbit following the $3\to6$ period-doubling, when $a$ increases.
When moving from $R_3$ to $R_4$, the stable real orbit looses its stability.
The intersection of the ``1'' line with $R_2$,
located  at $(a_*,b_*) \simeq (-0.94832, -0.94764)$,
is defined by algebraic numbers of degree $38$, a quite high degree.
This intersection has a {\it dual}\cite{eg2002} at
$b = 1/b_* \simeq -1.0552$ and $a\simeq-1.0560$.  
The tip of $R_2$, located roughly at $(a,b) =  (-0.95449,  -0.93348)$,
is defined by algebraic numbers of degree $76$.

All in all, the exact expressions reported here, 
result of long and elaborate algebraic computations,
reveal the existence of a new sort of bifurcation that
occurs among complex trajectories, transforming unstable ghost orbits into
stable complex orbits. Our Eqs~(\ref{eqorb}-\ref{eqeigen})
contain many additional features that will be considered in a 
future publication\cite{enga04}.

AE is a doctoral fellow of the Conselho Nacional de 
Desenvolvimento Cient\'{\i}fico e Tecnol\'ogico (CNPq), Brazil.
JACG is a Research Fellow of the CNPq.
This work was sponsored by CAPES (Brazil) and GRICES (Portugal).


\vspace{-0.5truecm}

\appendix

\section*{Appendix: The Coefficients $\theta_i(a,b;\sigma)$ and $\Theta_i(a,b)$}

\vspace{-0.5truecm}

The coefficients $\theta_i(a,b;\sigma)$ needed to define
{\it all period-$6$ orbits\/} are the following:
\begin{eqnarray*}
\theta_4 &=& [\sigma^2 + (1-b)\sigma  - 6a]/2, \\
\theta_3 &=& [2 \sigma^3 + 6(1-b)\sigma^2 -32a\sigma 
     -{\mathcal N}_1(1-b)/{\mathcal D}]/12,\\
\theta_2 &=& (1/24)\{ \sigma^4 +  6(1-b)\sigma^3 -
            [28a -3(1-b)^2]\sigma^2  - 6(1-b) \times \cr
         && [ 4a-(1+b^2)] \sigma
          + 36[ 2a - (1+b^2)] 
   -2(1-b)\sigma {\mathcal N }_1/ {\mathcal D} 
   + 3b {\mathcal N }_2/ {\mathcal D}   \},  \\
\theta_1 &=& (1/120)\{  \sigma^5 + 10(1-b)\sigma^4
             -5[ 8a -3(1-b)^2]\sigma^3   -10(1-b) \times \cr 
&& [10a -3(1+b^2)]\sigma^2
    + 12[ 22a^2 -13a(1+ b^2) -2b(1-b)^2]\sigma  \cr
&&   -(1-b)[ 5\sigma^2     + 5(1-b)\sigma -18a]  
        {\mathcal N }_1/ {\mathcal D} \cr
&&   + 3[5b\sigma-2(1-b^3){\mathcal N }_2/{\mathcal D} 
      -6b{\mathcal N }_3/ {\mathcal D}  \}, \\
\theta_0 &=& (1/720)\{ {\sigma}^{6}+15(1-b){\sigma}^{5}
             - 5[ 10a-9(1-b)^{2}] {\sigma}^{4} \cr
             &&-5(1-b)[16a-7{b}^{2}+2b-7]{\sigma}^{3}\cr
&&+ 2[272{a}^{2}-9(27b^2-10b+27)a  +9(5\,{b}^{2}-8b+5)(1-b)^{2}]
        {\sigma}^{2}\cr
&&+24(1-b)(15a-25{b}^{2}+26b-25)a\sigma -360{a}^{2}(2a-3(1+ {b}^{2} ))  \cr
&&-2 
   [(1-b)(5{\sigma}^{3}+15(1-b){\sigma}^{2} -44a\sigma) +15( {b}^{4} - b^3 \cr 
&&+ 6{b}^{2}- {b} +1)]{\mathcal N}_1 / {\mathcal D} 
     -9 [5b{\sigma}^{2}-(1-b)(4{b}^{2}-b+4)\sigma-10ba]
           {\mathcal N}_2 / {\mathcal D}  \cr 
&&+36b\sigma {\mathcal N}_3 /{\mathcal D}  
  -360b( 1+ {b}^{2} ){\mathcal N}_4 /{\mathcal D}
-5 (1-b)^{2} ( {\mathcal N}_1/{\mathcal D} )^2/2  \},
\end{eqnarray*}
where the following abbreviations are used:
\begin{eqnarray*}
{\mathcal D }   &=&   3{\sigma}^{4}-7(1-b)
    {\sigma}^{3}-(12a-11{b}^{2}-5b-11){\sigma}^{2} +(1-b)\times \cr 
&&(16a-19{b}^{2}-34b-19)\sigma 
-(1-b)^{2}(4a-8{b}^{2}-13b-8), \\
{\mathcal N }_1 &=&{\sigma}^{6}-2(1-b){\sigma}^{5}+
2(2a+5{b}^{2}-22b+5){\sigma}^{4} \cr
&&-4(1-b)(18a+5{b}^{2}-20b+5){\sigma}^{3}-
     [ 32{a}^{2}-8(22{b}^{2}-5b+22)a
\cr
&&+3(9{b}^{2}+4b+9)({b}^{2}+4b+1)]
{\sigma}^{2}+2(1-b)[64{a}^{2}-4(29{b}^{2}+46b\cr
&&+29)a + 51{b}^{4}+132{b}^{3}+258b^2+ 132b + 51]\sigma-12( 1-b
)^{2}\times \cr
&& [ 8{a}^{2}-(13{b}^{2}+28b+13)a
+({b}^{2}+b+1)(8{b}^{2}+13b+8)], \\
{\mathcal N }_2 &=&3{\sigma}^{6}-4(1-b){\sigma}^{5}-12(3a-4b
){\sigma}^{4} 
+2(1-b )(48a+5{b}^{2}\cr
&&-26b+5){\sigma}^{3}
       +[96{a}^{2}-24(7{b}^{2}+3b+7)a 
+13{b}^{4}+68{b}^{3}+126{b}^{2} \cr
&&+ 68b + 13 ]{\sigma}^{2}
-2(1-b) [64{a}^{2}-88(1+b)^{2}a
+ 27{b}^{4}+48{b}^{3}+90b^2\cr
&&+48b  +27]\sigma +4(1-b)^{2}(8a-9{b}^{2}-24b-9)a, \\
{\mathcal N }_3 &=& {\sigma}^{7}-2(10a-27b){\sigma}^{5} 
+2(1-b)(20a-7{b}^{2}-55b-7){\sigma}^{4}+ [64{a}^{2}\cr
&&-16(8{b}^{2}+11b+8)a
+(7{b}^{2}+4b+7)(9{b}^{2}+10b+9)]{\sigma}^{3} \cr 
&&+2(1-b)[16{a}^{2}+4(29{b}^{2}+53b+29)a
-(105{b}^{4}+211{b}^{3}+376b^2 \cr 
&&+ 211b +105)]{\sigma}^{2}
-4(1-b)^{2} [48{a}^{2}-(25{b}^{2}+42b+25)a -(8{b}^{2}\cr
&&+13b+8)(7{b}^{2}+4b+7)]\sigma+24(1-b)^{3}(4a-8{b}^{2}-13b-8)a,
  \\
{\mathcal N }_4 &=& {\sigma}^{6}-(1-b){\sigma}^{5}-(14a+{b}^{2}-17b
+1){\sigma}^{4}
+(1-b)(34a-3{b}^{2}\cr
&&-43b-3){\sigma}^{3}
+[40{a}^{2}-2(29{b}^{2}-b+29)a 
-4{b}^{4}+27{b}^{3} +8b^2 \cr 
&&+ 27b -4]{\sigma}^{2}
      -(1-b) [96{a}^{2}-2(69{b}^{2}+110b+69)a\cr
&&+16{b}^{4}+87{b}^{3}+178b^2 + 87b+16]\sigma +2(1-b)^{2}\times \cr
&&[28{a}^{2}-9(6{b}^{2}+11b+6)a+3({b}^{2}+b+1)(8{b}^{2}+13b+8)],\\
   &&\cr
{\mathcal N }_5 &=& {\sigma}^{8}
      +2\,(-9\,a+2\,{b}^{2}+17\,b+2){\sigma}^{6}
      +2\,(b-1)(-6\,a+9\,{b}^{2}+29\,b \cr
  && +9){\sigma}^{5} 
     +(72\,{a}^{2}-12\,(3\,{b}^{2}+22\,b+3)a
     +186\,{b}^{2}+82\,b+59\,{b}^{4}     \cr
  && +82\,{b}^{3} +59){\sigma}^{4}  +2\,(b-1)
     [8\,{a}^{2}-12\,(3\,{b}^{2}+16\,b+3)a  \cr
  && +({b}^{2}+b+1)(111\,{b}^{2}+116\,b+111) ]{\sigma}^{3}
     +2[-32\,{a}^{3}-24\,(2\,{b}^{2}  \cr
  && -9\,b +2){a}^{2} +(115\,{b}^{4}-56\,{b}^{3}
      -226\,{b}^{2}-56\,b+ 115 )a         +104\,{b}^{6}\cr  
  && +36\,{b}^{5} +22\,{b}^{4}+22b^2+36\,b+104]{\sigma}^{2}
      +4\, (b-1 ) [-64\,{a}^{3}\cr
  && +4\,(21\,{b}^{2}+62\,b+21){a}^{2}
        -(11\,{b}^{4}+196\,{b}^{3}+402\,{b}^{2}+196\,b+11)a\cr
  &&+ ({b}^{2}+b+1) (16\,{b}^{4}+87\,{b}^{3}
           +178\,{b}^{2}+87\,b+16) ]\sigma\cr
  &&+8(b-1)^{2}[-24\,{a}^{3}+5\,(11\,{b}^{2}+20\,
        b+11){a}^{2}-9\,(6\,{b}^{2}+11\,b  \cr
 &&       +6)({b}^{2}+b+1)a
     +3\,(8\,{b}^{2}+13\,b+8)({b}^{2}+b+1)^{2}].
\end{eqnarray*}


The coefficients $\Theta_i = \Theta_i(a,b)$ needed in
Eq.~\ref{eqsigma} to obtain the
$9$ solutions $\sigma_\ell = \sigma_\ell(a,b)$ are 
$\Theta_9=  1$,\quad
$\Theta_8= 1-b$,\quad
$\Theta_7 = -24a+2({b}^{2}+16b+1)$,\
and
\begin{eqnarray*}
\Theta_6&=& 2(1-b)[4a-(7{b}^{2}+12b+7)],\\
\Theta_5&=& 144{a}^{2}-16({b}^{2}+25b+1)a+  
(49{b}^{4}+52{b}^{3}+266{b}^{2}+52b+49),\\
\Theta_4&=&  -(1-b) [ 112{a}^{2}
-16({b}^{2}+27b+1)a \cr 
&&+(175{b}^{4}+388{b}^{3}+518{b}^{2}+388b+175)], \\
\Theta_3&=& -4[ 64{a}^{3}
-8(b+5)(5b+1){a}^{2} 
-2(17{b}^{4}-48{b}^{3}-172{b}^{2}\cr
&&-48b+17)a 
-(7{b}^{2}+2b+7)(5{b}^{4}+9{b}^{3}-{b}^{2}+9b+5)],\\
\Theta_2&=& 4(1-b) [ 64{a}^{3}
             -8(15{b}^{2}+38b+15){a}^{2} 
             +6(23{b}^{4}+48{b}^{3}+96{b}^{2}\cr
&&+48b+23)a 
       -(7{b}^{4}+33{b}^{3}+55{b}^{2}+33b+7)(7{b}^{2}+2b+7)], \\
\Theta_1&=&8(1-b)^{2} [ 32{a}^{3}
            -2(19{b}^{2}+34b+19){a}^{2}
             -2(26{b}^{4}+23{b}^{3}+9{b}^{2}\cr 
&&+23b+26)a 
        +({b}^{2}+b+1)(7{b}^{2}+2b+7)(8{b}^{2}+13b+8)], 
  \\
\Theta_0&=& -16a(1-b)^{3}[16{a}^{2}-(37{b}^{2}+62b+37 )a
            +3({b}^{2}+b+1)        (8{b}^{2}+13b+8)].
\end{eqnarray*}

\end{document}